\documentclass[aps,showpacs,prl,twocolumn]{revtex4-2}  

\usepackage{amssymb}
\usepackage{bm}
\usepackage{amsmath}
\usepackage{graphicx}
\usepackage{hyperref}
\usepackage{xcolor}
\usepackage{ulem}

\newcommand{\lidar}{lidar}

\begin{document}

%
\title{Demonstration of quantum-enhanced rangefinding robust against classical jamming}

\author{M. P. Mrozowski}
\author{R. J. Murchie}
\author{J. Jeffers}
\author{J. D. Pritchard}
\email{jonathan.pritchard@strath.ac.uk}

\affiliation{Department of Physics and SUPA,University of Strathclyde, Glasgow G4 0NG, United Kingdom}

%

\begin{abstract} 
In this paper we demonstrate operation of a quantum-enhanced lidar based on a continuously pumped photon pair source combined with simple detection in regimes with over 5 orders of magnitude separation between signal and background levels and target reflectivity down to -52 dB.  We characterise the performance of our detector using a log-likelihood analysis framework, and crucially demonstrate the robustness of our system to fast and slow classical jamming, introducing a new protocol to implement dynamic background tracking to eliminate the impact of slow background changes whilst maintaining immunity to high frequency fluctuations. Finally, we extend this system to the regime of rangefinding in the presence of classical jamming to locate a target with an 11 cm spatial resolution limited only by the detector jitter. These results demonstrate the advantage of exploiting quantum correlations for lidar applications, providing a clear route to implementation of this system in real-world scenarios.
\end{abstract}
\maketitle



Optical \lidar{} is a pivotal technology for achieving precise target detection and rangefinding with high spatial precision \cite{mcmanamon12,nanxi22}, utilised in a range of applications from performing ground surveys \cite{amann01}, monitoring sea levels \cite{hooijer21}, to aiding navigation in autonomous vehicles \cite{royo19}. Under conditions necessitating low-light levels and a substantial background, arising from low target reflectivity, environmental noise, or deliberate jamming, classical \lidar{} techniques fail to discern between signal and noise photons leading to diminished signal-to-noise ratio and an inability to detect targets confidently. 

Significant progress has been made towards exploiting \lidar{} at the single photon level \cite{massa98} enabled by advances in detector technologies and computational analysis to enable 3D imaging using single-pixel detection \cite{sun16} or single photon cameras \cite{tachella19} suitable for operating in adverse backgrounds \cite{wallace20,tobin21}, however typically these devices operate using strong modulated classical light sources for target illumination to compensate for low return probability. In contrast, quantum-enhanced illumination \cite{pirandola18} offers a compelling alternative approach, whereby the utilization of non-classical heralded photon sources affords the exploitation of coincidence detection techniques, enabling effective background photon suppression without temporal modulation of the signal source \cite{england19,liu19,liu20}.

The original framework for quantum illumination proposed by Lloyd \cite{lloyd08} demonstrated that by exploiting entanglement it was possible to out-perform classical systems, with an extension to Gaussian state analysis bounding the maximum quantum advantage to 6~dB assuming an unknown optimal measurement \cite{tan08,shapiro20}. Measurement protocols have been proposed offering up to 3~dB advantage \cite{guha09,zhang15}, whilst a detection scheme able to exploit the full quantum advantage \cite{zhuang17,zhuang17a} remains a significant technical challenge. Experimental demonstrations of quantum illumination with phase-sensitive detection have been performed in-fiber with noise added at the detectors \cite{zhang15}, with recent extensions to operation in the microwave domain compatible with radar applications \cite{luong19,chang19,barzanjeh20}.

\begin{figure*}[t!]
  \centering
  \includegraphics[width=\textwidth]{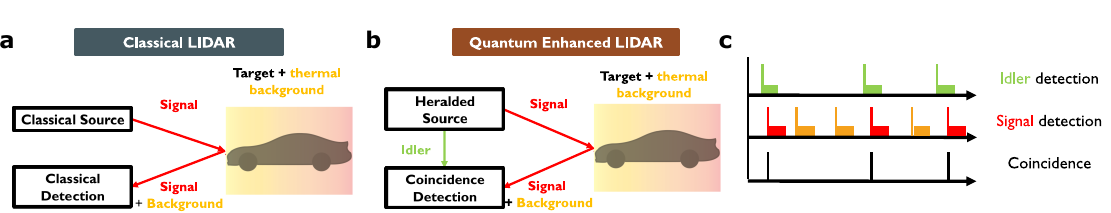}
  \caption{
  An overview of quantum enhanced \lidar{} based on simple detection. 
  \textbf{a} A weak classical light source illuminates a weakly reflecting target, which is embedded within a noisy background environment. 
  A detector is unable to distinguish between reflected signal photons or background photons, leading to suppressed signal to noise and increased time required for confident target detection. 
  \textbf{b} In quantum-enhanced \lidar{} the source is replaced with a non-classical photon pair source. 
  One photon from each pair is detected locally (idler) and the signal is sent to probe the target. 
  Strong temporal correlation between signal and idler can be exploited to perform coincident detection (shown in \textbf{c}) to both eliminate counts due to background (thereby enhancing the SNR) and providing distance information from the delay at which a temporal correlation is observed.}
  \label{fig:q\lidar{}}
\end{figure*}

A much simpler approach is to exploit the temporal correlations arising from photon pairs generated using spontaneous parametric downconversion using either pulsed or continuous (CW) sources \cite{yang21}. Experimental demonstrations have shown enhancement in the signal-to-noise ratio by exploiting these correlations \cite{lopaeva13,balaji18,england19,liu19}, offering robust operation with respect to classical jamming \cite{england19,liu20} and first demonstrations of rangefinding \cite{rarity90,liu19,frick20}. 
Recently, using dispersion compensating fibers a further enhancement in quantum detection was demonstrated by spreading the noise across multiple time-bins whilst preserving the temporal correlation of the two-photon coincidence to achieve 43~dB times enhancement in SNR at noise levels up to three times the signal level \cite{blakey22}. Additionally, detector multiplexing \cite{yang22} permits multi-mode range-finding that can enable covert operation using light that is spectrally and statistically indistinguishable from the background \cite{frick20}.

In this paper we present experimental demonstration of quantum enhanced \lidar{} using a CW source of heralded photon pairs with simple detection. Using the log-likelihood analysis framework \cite{murchie21,murchie23} we characterise the performance of our detector, demonstrating operation of the quantum \lidar{} in regimes with over 5 orders of magnitude separation between signal and background levels and target reflectivity down to -52 dB. This corresponds to up to 30 dB enhancement over classical illumination, or 17 times faster target discrimination to achieve a comparable error rate. Using this system, we demonstrate the robustness of our quantum enhanced \lidar{} approach to classical jamming, implementing a dynamic analysis procedure to track slow variations in background noise and immunity to high frequency fluctuations. Finally, we extend this system to perform rangefinding, implementing moving target discrimination with a 11~cm spatial resolution, limited by the timing jitter of our detectors. These results demonstrate the advantage of exploiting quantum correlations for \lidar{} applications, and provide a clear route to realistic use of this system in a scenario comparable with real-world operations.


\section{Results}
\subsection{Quantum Enhanced \lidar{}}

A schematic illustration of classical \lidar{} is shown in Fig.~\ref{fig:q\lidar{}}$\textbf{a}$ with a classical light source sending light towards a target, and a detector used to collect the return signal reflected from the target. The target is embedded within a noisy background, which here we treat as Poissonian but could be described by any underlying mode distribution such as thermal light. In the limit where the noise level is large compared to the returning signal (either due to using a weak source for covert operation or due to a target at large distance or with low reflectivity) this leads to a significant reduction in signal to noise as it is not possible to discriminate between background and signal photons at the detector, increasing the time required to reach a confident detection. In this setup, a continuous source can be used to determine if there is a target, but to obtain information regarding range it is also necessary to introduce temporal modulation of the source, for example pulsed operation or amplitude modulation. In a low noise regime this modulation reduces the covertness of the illumination source, whilst an uncooporative target can easily jam or spoof the detector by sending an artificially delayed or modulated signal.


In the quantum enhanced \lidar{} scheme shown in Fig.~\ref{fig:q\lidar{}}$\textbf{b}$ the light source is replaced by a non-classical of photon pairs which are generated via spontaneous parametric down conversion (SPDC) leading to pairs being created at random times. One photon from each pair is detected locally (the idler), whilst the signal photon is directed at the target, enabling use of coincidence detection to discriminate between signal detector counts due to background photons and those arising from signal reflection (as illustrated in Fig.~\ref{fig:q\lidar{}}c). Crucially in this regime we are not exploiting entanglement, but instead the strong temporal correlations intrinsic to the pair generation process to suppress background counts, offering a significant improvement in SNR and a robustness to jamming that we will demonstrate below. Additionally, the coincident detection permits range-finding without temporal modulation of the laser, making it much harder to spoof or intercept than classical techniques.

\begin{figure*}[t!]
  \centering
  \includegraphics[width=\textwidth]{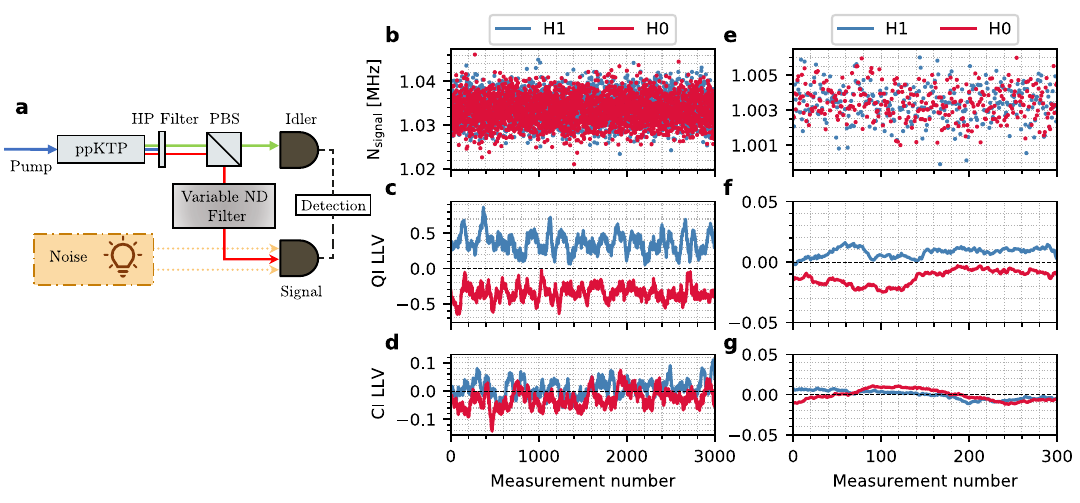}
  \caption{
  \textbf{a} Experimental setup used to conduct the experiment, 
  here the ppKTP crystal is pumped with 405 nm CW light to generate an 810 nm photon pair in the process of SPDC Type-II, 
  polarising beam splitter (PBS) is used to separate signal \& idler pair. 
  The idler photons are detected locally, while the signal photons experience loss passing though a neutral density (ND) filter before detection. 
  Together with signal, noise is injected into the signal detector.
  \textbf{b-d} Measurement in which signal is experiencing 33.5 dB loss in presence of 1 MHz background noise corresponding to SNR$_{\text{CI}}$ of -37.9 dB. 
  In \textbf{b} detection events registered by the signal detector are shown, while \textbf{c} and \textbf{d} show LLV analysis applied to coincidence detection and classical detection respectively.
  In \textbf{e-g} we repeat the measurement after increasing the loss to 52 dB and reducing the SNR$_{\text{CI}}$ to -51.5 dB, here to compensate for the low signal return of 7.1 $\pm$ 0.9 Hz we increase the integration time per measurement to 1 s. Following the established convention we show signal detector countrate in \textbf{e}, LLV analysis with 150 measurement moving average applied to quantum enhanced \textbf{f} and classical \textbf{g} data.
  }
  \label{fig:llv}
\end{figure*}

\subsection{Log-Likelihood Framework} 
In both target detection and range finding the aim is to evaluate, based on a finite set of measurements, whether or not a target is present. This reduces to a state discrimination problem of deciding whether the measured statistics correspond to the target being present (hypothesis 1 - H1) or absent (hypothesis 0 - H0). Prior analysis of this problem has focused on the fundamental bounds given an unknown optimal measurement scheme \cite{lloyd08,tan08} or formulated this as a probe transmission estimation problem whereby the uncertainty of the transmission estimation is described by the Cram\'{e}r–Rao bound \cite{liu19,liu20}. In either approach, a fundamental issue lies in defining the threshold at which to consider a target present or absent.

In this work we use an alternative analysis protocol based on the log-likelihood value (LLV) $\Lambda$ defined as \cite{murchie21,murchie23}
\begin{equation}
\Lambda(x,k) = \ln\left(\frac{P_{\mathrm{H1}}(x,k)}{P_{\mathrm{H0}}(x,k)}\right),\label{eq:LLV}
\end{equation}
where $x$ represents the measured detector count data, $k$ is the number of trials and $P_{\mathrm{H{1,0}}}(x,k)$ represents the the probability that the target is there (or not) given $x$ clicks after $k$ trials. When calculating $\Lambda$ in the case of quantum illumination (QI) $x$ is the number of detected coincidence events in a measurement time $T$ for a given coincidence window $\tau_c$, and $k$ is the number of idler clicks, whilst for classical illumination (CI) $x$ is the number of signal detection events and $k=T/\tau_c$.

An advantage of using the log-likelihood ratio is that it gives a natural threshold value of $\Lambda=0$ that is independent of the system parameters, providing a degree of self calibration in the system. If $\Lambda<0$ it is more likely that the target is absent, whilst for $\Lambda>0$ it is more likely that a target is present. To evaluate the error associated with making a decision on wheter a target is present using this threshold, we introduce the distinguishability $\phi = 1-\left[(1-P_\mathrm{D})+P_\mathrm{FA}\right]$ where $P_\mathrm{D}$ is the probability of correctly detecting a target that is there and $P_\mathrm{FA}$ is the false alarm probability caused by incorrectly detecting a target when there is not one present. Analytically these probabilities are evaluated by integrating over the underlying LLV probability distributions associated with H1 and H0 respectively, whilst experimentally we evaluate distinguishability by calculating the fraction of data points with $\Lambda>0$ when the object is present or absent. 

\subsection{Quantum-Enhanced Detection}
The accurate identification of target objects probed with \lidar{} systems is challenged by loss and noise. Loss encompasses attenuation due to absorption, scattering, reflection, and transmission, leading to weakened signal return. Noise involves unwanted signals from detectors (dark counts), environmental factors (thermal background, jamming), and system imperfections, introducing inaccuracies and distortions. Addressing these challenges often requires advanced signal processing, sensor design, and calibration techniques \cite{lidar-challenges}. 
Here we demonstrate the performance of our simple \lidar{} system using single photon counting modules based on single photon avalanche diodes (SPADs) for detection.

A schematic of the setup used to conduct the experiment is shown in Fig.~\ref{fig:llv}\textbf{a}. Here a photon pair source generates photons at 810~nm using type-II SPDC from 405~nm CW pump in a ppKTP crystal. Details of the source characterisation can be found in Methods section. During the experimental trials the resulting photon pair is separated on a polarising beam splitter and the idler photon is detected directly.  

Noise is injected to the system using an 810~nm LED driven by a low-noise current driver to provide a constant noise independent of target reflectivity or position. This noise is combined with the attenuated signal and coupled into a single photon detector. To emulate controllably the loss associated with finite target reflectivity in the object present (H1) case, the signal photons pass  through a calibrated neutral density filter. For the target absent case (H0) a beam block is placed in the signal path, allowing only the noise to be coupled into the signal detector. 

Fig.~\ref{fig:llv} shows the performance of the \lidar{} system for detecting a stationary target operating under two distinct loss regimes of 33.5 and 52~dB, approaching values typically encountered in real \lidar{} systems ($\geq$ - 50 dB) \cite{frick20}. In both cases, the system operates in a regime with an average background count rate of 1~MHz, and using a coincidence window of $\tau_c = 2$~ns.


For the data shown in Fig.~\ref{fig:llv}\textbf{b-d} with a loss of 33.5~dB, the crystal is pumped at 50~$\mu$W giving a pair production rate of $377\pm5$ kHz. For the object present case, this gives an effective signal return rate of $167\pm1$~Hz, corresponding to a classical signal to noise of $\mathrm{SNR}_\mathrm{CI}=-37.9\pm0.1$~dB. To evaluate the quantum SNR ratio, we take the ratio of the measured coincidence rate with target present and noise source turned off, against the number of accidental coincidences recorded with the noise source enabled and target absent. For this data we find values of $39.1\pm0.4$~Hz and $200.2\pm0.5$~Hz respectively giving $\mathrm{SNR}_\mathrm{QI}=-7.1\pm0.1$~dB. We acquire 3050 consecutive measurements using an integration time of $T=0.1$~s for both target present and target absent scenarios. Figure~\ref{fig:llv}\textbf{b} shows the raw signal counts measured in each case, showing that the additional signal counts with target present are indistinguishable compared to the $\sim1000$ count standard deviation of the background counts. 

\begin{figure}[t!]
  \centering
  \includegraphics[width=\columnwidth]{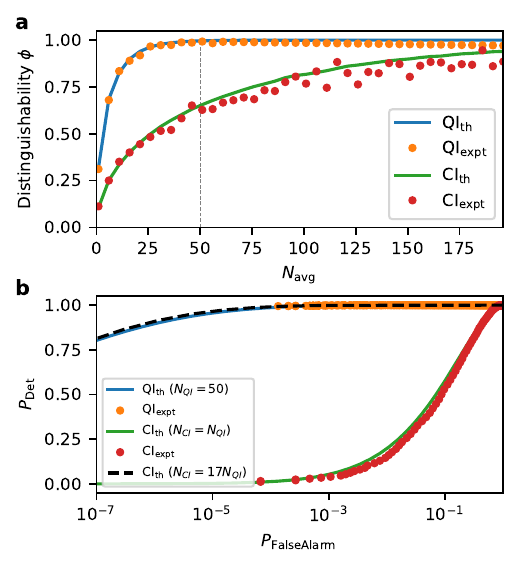}
  \caption{Characterising system performance for data presented in Fig.~\ref{fig:llv}$\textbf{b-d}$. $\textbf{a}$ shows the experimental and theoretical distinguishabilities $\phi$ as a function of number of averages $N_\mathrm{av}$ for both CI and QI data and theory, with a peak value of $\phi_\mathrm{QI}=0.995\pm0.003$ for an integration window of 50 (corresponding to 5~s). $\textbf{b}$ shows the experimental and theoretical receiver-operator curve (ROC) for the optimum average window length of $N_\mathrm{av}=50$. For comparison, CI requires a $17\times$ longer averaging time to achieve an equivalent performance to QI as indicated by the dashed line. The discrepancy between CI theory and data is due to the sensitivity to small drifts in background noise level across the measurement window.\label{fig:ROC}}
\end{figure}

To apply the LLV analysis to the data, the single shot probability distributions $p_{\text{H}_{1,0}}$ are estimated using the first 50 measurements of each case (see Methods). From this, the LLV $\Lambda(x,k)$ for each data point can be calculated, resulting in single shot distinguishabilities of $\phi_\mathrm{QI}=0.31\pm0.01$ and $\phi_\mathrm{CI}=0.086\pm0.003$ respectively, demonstrating the enhancement in detection performance using QI. To enhance further the distinguishability we perform a rolling window average with $N_\mathrm{av}=50$ to smooth the data. Figure~\ref{fig:llv}\textbf{c-d} shows the corresponding averaged LLV data for QI and CI, which clearly reveals that despite the relatively small change in signal level on the detector the QI enhanced \lidar{} is able to discriminate between target present and target absent robustly, with the corresponding averaged $\Lambda$ values never crossing zero. The classical LLV however is significantly noisier, with both cases frequently crossing the detection threshold leading to significant error if using this for discrimination.

\begin{figure*}[ht!]
  \centering
  \includegraphics{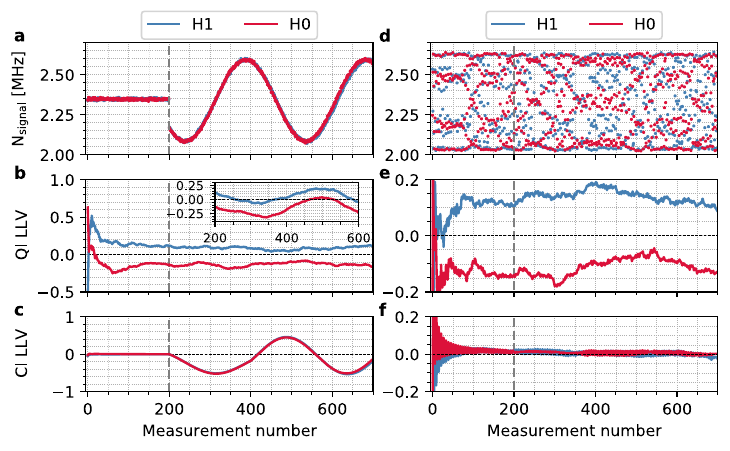}
  \caption{
  Two jamming experiments, 
  in \textbf{a-c} we apply slow modulation after estimating the H1 \& H0 probabilities with static background for 200 measurements (grey vertical dashed line) 
  and in \textbf{d-f} we apply fast modulation of the background and estimate the probabilities with dynamic background (grey vertical dashed line). 
  Both experiments were performed under identical conditions of 33.5 dB loss, 2.3 MHz average background, 0.3 MHz modulation amplitude and integration time of 0.1 s. 
  Here in \textbf{a} and \textbf{d} signal detector count rates are plotted. In \textbf{b} and \textbf{e}  LLV analysis applied to our quantum enhanced experimental data is shown, here on plot \textbf{b} the inset shows the LLV analysis without dynamic background tracking, and the main plot shows the results with us dynamically tracking the background and correcting for the accidental coincidence counts resulting from increasing/decreasing the background.
  In \textbf{c} and \textbf{f} results of classical LLV analysis were plotted.
  This results show that quantum enhanced \lidar{} is robust to classical jamming.
  }
  \label{fig:jamming}
\end{figure*}

In Fig.~\ref{fig:llv}\textbf{e-g} the loss is increased to 52~dB, with the crystal pump power increased to 150~$\mu$W to increase the pair rate to $1.13\pm0.02$ MHz whilst maintaining the same level of background noise.
In this regime the effective signal return rate in the object present case reduces to $7.1\pm0.9$~Hz, corresponding to a classical signal to noise of $\mathrm{SNR}_\mathrm{CI}=-51.5\pm0.6$~dB. 
The coincidence counts rates with target present and noise disabled equal to  $1.8\pm0.1$~Hz and for target absent with noise engaged equal to $577\pm1$ Hz respectively, leading $\mathrm{SNR_{QI}}=-25.1\pm0.2$.
As the system is now operating with a signal over 5 orders of magnitude smaller than the background noise, the integration time is increased to $T=1$~s for these measurements. 
As before, Fig.~\ref{fig:llv}\textbf{e} shows that the raw signal counts are indistinguishable, however using a rolling window of $N_\mathrm{av}=150$ from Fig.~\ref{fig:llv}\textbf{f,g} it is clear that the QI LLV is able to discriminate between target present and absent cases despite these challenging operating parameters whilst the CI LLV is entirely unreliable, with distinguisbabilities of $\phi_\mathrm{QI}=0.67\pm0.22$ and $\phi_\mathrm{CI}=0.33\pm0.02$ for the window-averaged data.

To characterise system performance, we perform further analysis of the data presented in Fig.~\ref{fig:llv}\textbf{b-d} for 31.5~dB loss. In Fig.~\ref{fig:ROC}\textbf{a} the experimental and analytical distinguishability are plotted as a function of average window size $N_\mathrm{av}$. These data reveal the for $N_\mathrm{av}=50$ (indicated by the dotted line) the quantum-enhanced \lidar{} reaches a maximum distinguishability $\phi_{\mathrm{QI}} =  0.995\pm0.003$ in excellent agreement with theory, whilst the classical system only achieves $\phi_{\mathrm{CI}} =  0.63\pm0.02$. Extending to longer integration times we observe the data for both CI and QI begins to deviate from the theoretical predictions due to long term drift in both the noise source and pump laser intensities.

Using this optimal average window size, in Fig.~\ref{fig:ROC}\textbf{b} we present the experimental and theoretical receiver operator curve (ROC) obtained by scanning the threshold value of $\Lambda$ at which a target is deemed to be present. For the QI data we obtain excellent agreement with theory, and at $\Lambda=0$ this corresponds to a false-alarm probability of $P^\mathrm{QI}_\mathrm{FA}=5\times10^{-4}$, whilst for CI $P^\mathrm{QI}_\mathrm{FA}=0.25$. To quantify the performance advantage offered by the quantum enhanced \lidar{} in this regime we analytically find the number of averages required for the CI ROC curve to match the observed QI data, shown as a black dashed line on the Figure. This analysis shows that for equivalent performance the classical detector would need to integrate 17$\times$ more data, demonstrating a significant performance advantage when operating in regimes of significant loss and high background.

\subsection{Resilience to jamming}
Classical jamming of \lidar{} systems refers to intentional interference aimed at disrupting the operation of \lidar{} technology, as well as operation in environment where level of background noise fluctuates. Intentional jamming techniques involve emitting strong modulated light or laser signals, or deploying countermeasures to confuse or overwhelm the \lidar{} sensor. The objective of intentional jamming of \lidar{} systems is to hinder accurate data gathering, compromise situational awareness, or impede target detection and classification.
Such jamming activities can lead to impaired perception and navigation capabilities in autonomous vehicles and other \lidar{}-dependent applications, potentially resulting in hazardous scenarios. In the following we demonstrate the resilience of quantum enhanced \lidar{} to dynamic jamming using both slow and fast modulation of the background noise. For both of these experiments the target loss was set to 33.5~dB, pump power to 50~$\mu$W, $T=0.1$~s, with an average background of 2.3~MHz modulated with an amplitude of 0.3~MHz.
 
Figure~\ref{fig:jamming}\textbf{a-c} shows the effect of applying a slow sinusoidal background modulation after an initial period of constant background which is used to estimate the single shot probabilities $p_{\text{H}_{1,0}}$. In the inset of Fig.~\ref{fig:jamming}\textbf{b} we show the QI LLV evaluated assuming the initial constant background data, which shows that whilst we maintain a clear separation between the QI LLV for object present and absent case, the modulation is visible in the data.

To mitigate this effect we implement dynamic background tracking by using the raw-signal data to estimate the average background level associated with each measurement (valid in this regime where $\bar{n}_\mathrm{bg}\gg \bar{n}_\mathrm{sig}$). Using our model (see Methods), we create a look-up table (LUT) of probabilities $P_{\text{H}_{1,0}}$ for different background levels $\bar{n}_\mathrm{bg}$ whilst keeping all other parameters constant. For each measurement we then use the raw signal counts to assign the appropriate probability distribution when calculating the single shot LLV. This pre-calculated LUT approach can be used to enable real-time implementation in future experiments using the signal count-rate to track the background. The resulting QI LLV is shown in the main plot of Fig.~\ref{fig:jamming}\textbf{b}, which has now eliminated the modulation and shows the QI \lidar{} can be made immune to slow jamming with a single shot distinguishability of $\phi_\mathrm{QI}=0.15\pm0.03$ despite the 26~\% background modulation. For comparison, Fig.~\ref{fig:jamming}\textbf{c} shows the classical LLV (for which no background correction is possible) is completely unable to distinguish between the two regimes and has been spoofed by the jamming signal. 

In Fig.~\ref{fig:jamming}\textbf{d-f} we perform a second experiment where a fast white noise source is now added to the slow classical modulation, resulting in a pseudo-random noise level seen from the signal on the number of signal counts in~\textbf{d}. As with the slow modulation, the QI LLV is immune to the fast noise whilst the classical LLV data is entirely washed out with the fast noise causing the CI LLV to average to zero making it unable to distinguish if a target is present. These data highlight the advantage of QI not only in performing faster discrimination for a given average signal level, but also in providing a system robust to jamming.

\begin{figure}[t!]
  \centering
  \includegraphics[width=\columnwidth]{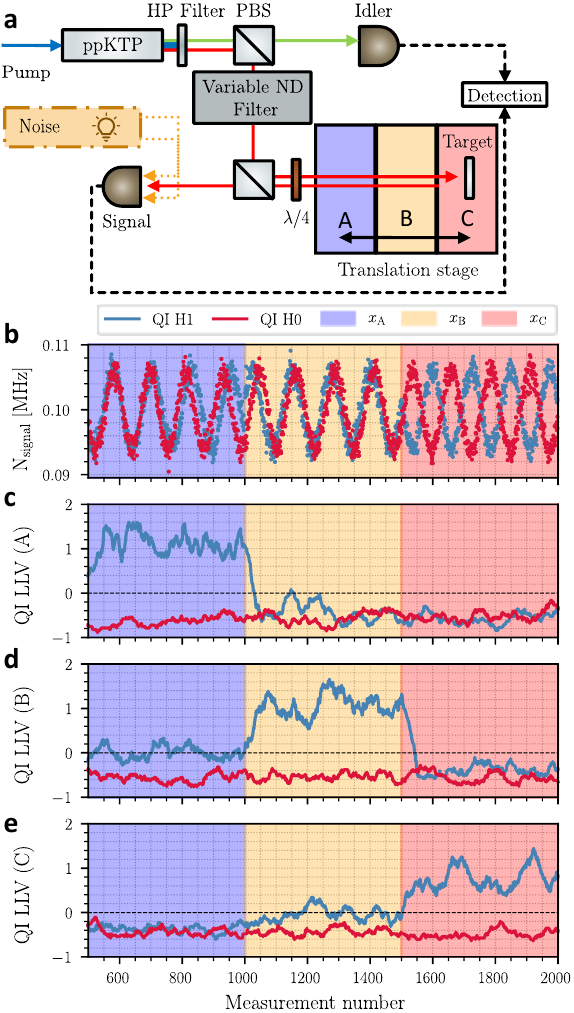}
  \caption{Quantum-Enhanced Rangefinding. We apply a modulated background noise and translate the target from an initial location of $x_A=0$ to $x_B=11$~cm after 1000 measurements, moving to $x_C=22$~cm after 1500 measurements. $\textbf{a}$ Raw signal counts showing noise modulation is constant independent of target location. $\textbf{b-d}$ QI LLV with coincidence channel delays set to probe targets at $x_{A,B,C}$ respectively. These results show the QI can perform confident range finding even in the presence of a significant background.} 
  \label{fig:rangefinding}
\end{figure}

\subsection{Rangefinding}
The main application of \lidar{} based systems involves use of time-of-flight detection to estimate the distance to targets. In this section we extend our quantum enhanced \lidar{} to demonstrate active rangefinding even in the presence of classical noise.

The modified experimental setup used for rangefinding is shown in Fig.~\ref{fig:rangefinding}\textbf{a}.
In order to simulate variations in target position, a mirror serving as the target was positioned on a motorized translation stage enabling the target to be moved a total range of 22~cm. We define three locations separated by 11 cm intervals, denoted as $x_A$, $x_B$, and $x_C$, and assign three parallel coincidence detection channels with delays of $\tau_{x_A}$~=~1.77~ns, $\tau_{x_B}$~=~2.52~ns, and $\tau_{x_C}$~=~3.27~ns corresponding to the round-trip time to each location. As above, the experiment was performed with loss of 33.5~dB and 50~$\mu$W pump power, but with an average background of 0.1 MHz background and jamming amplitude of 10 kHz. To achieve better resolution and mitigate the cross-talk between target positions due to the 250~ps jitter of our detectors, the coincidence window $\tau_c$ was set to 0.2~ns.

During the experiment we acquire data for each coincidence channel in parallel, with target moved from an initial position of $x_A$ to $x_B$ after 1000 measurements and finally to $x_C$ after 1500 measurements. The results are shown in Fig.~\ref{fig:rangefinding}\textbf{b-e}, where as before the raw signal counts are indistinguishable for both target presence and position whilst the corresponding QI LLV channels calculated for $N_\mathrm{av}=50$ clearly reveal the quantum-enhanced \lidar{} is able to resolve robustly the location of and hence track the target as it moves between the three distinct locations, despite there being no active modulation of the source, and in the presence of intentional classical noise. 

Note that since the time-reference for range finding comes exclusively from the heralding of the idler photon, the CI detection method is unable to provide any rangefinding information. Further, the SPDC generation of photon pairs from a CW pump provides robustness against spoofing due to the covertness of illumination, which appears to the target as a weak thermal source within the environmental background.

\section{Discussion}

In conclusion, we have presented experimental demonstration of a quantum-enhanced \lidar{} system utilising a log-likelihood analysis framework for target detection and rangefinding, robust to classical jamming and dynamic background changes in challenging regimes of high background and low signal rates. This work significantly expands beyond prior research on quantum \lidar{} based on correlated photon sources\cite{england19,frick20,liu19,liu20,rarity90,blakey22}, specifically by operating in the regime of both large environmental backgrounds (SNR~$<$~-50 dB) and low signal returns ($>$~50~dB attenuation), compatible with realistic \lidar{}\cite{frick20}. Our observations reveal a significant improvement in the signal-to-noise ratio of up to 30~dB when comparing classical and quantum \lidar{}, with the ultimate limitation in quantum gain related to the second order correlation of the pair source $g_{s,i}^2(0)$ \cite{england19}. This improvement represents an approximate 13 dB increase in SNR compared to the findings in \cite{england19}, and is comparable to the results reported in \cite{liu19}. In this regime we show QI can provide high fidelity target discrimination on timescales 17$\times$ faster than classical techniques. In future, further enhancements can be achieved by employing more sophisticated detection techniques such as dispersion compensation tailored to pulse sources\cite{blakey22}.

Our quantum-enhanced \lidar{} approach is resilient against both sinusoidal and white-noise modulated classical jamming, and we present a new protocol for active background tracking to reduce sensitivity to slow drifts or intentional spoofing attempts, while remaining immune to high-frequency fluctuations. Specifically, this technique works optimally in the regime of high background where the instantaneous count-rate measured on the signal detector provides a real-time probe of the background rate which can then be used to improve the resilience against noise when analysing small changes in the measured coincidence rate.

Applying these techniques in a range-finding modality, we demonstrate the ability not only to perform target detection but also determine the location of the target in the presence of active jamming. Currently we demonstrate a resolution of 11~cm, limited only by the timing jitter of the room temperature SPADs motivated by the ability to incorporate these into a low power, portable device. Enhanced performance is possible using superconducting nanowire detectors offering significantly reduced timing uncertainty \cite{Taylor20,McCarthy13,Buller11} at the cost of requiring a cryogenic cooling system.

While the ability to achieve centimeter-level resolutions with room temperature SPADs is promising for practical applications, it is important to acknowledge the fundamental limitation that long-range or low-reflectivity targets require a sufficient photon flux for at least one return photon within the experiment window. This makes detection of uncooperative targets at distance challenging, with the demonstrated -52~dB loss comparable to that expected from a Lambertian scatterer at 15~m using a 10~cm diameter telescope. Nonetheless, our results demonstrate that quantum-enhanced \lidar{} offers a practical speedup in time to detection, allowing for operation at lower light levels, while presenting a low-intensity random thermal signal to observers. We also note that the detection of cooperative targets could, even with the results presented here, be accomplished at significant distances.

Compared to classical \lidar{} methods that rely on amplitude-modulated pulses for rangefinding, our use of continuous-wave sources mitigates the risk of being spotted and spoofed by observers as peak pulse intensity increases. These results emphasize the advantages of exploiting quantum correlations for \lidar{} applications and provide a clear pathway towards the realistic deployment of this system in scenarios comparable to real-world operations.
Overall, our research contributes to the advancement of quantum-enhanced \lidar{} technology, paving the way for its integration into practical applications where enhanced performance and security are crucial factors.

\section{Methods}
\subsection{Experiment Setup}
\subsubsection{SPDC Photon Source}\label{Methods:SPDC}

The QI source used in these experiments is a type-II collinear SPDC photon pair source based on a ppKTP crystal \cite{lee16}, with a polling period of 
$
10~\mu$m. 
The crystal was pumped using a continuous wave 405~nm pump laser focused to a $1/e^2$ radius of 11 $\mu$m to generate photon pairs at 810~nm. 
The source was characterised using a Hanbury-Brown-Twiss configuration \cite{beck07}. 
At a pump power $P = 0.3~$mW, the typical count rates were $N_\mathrm{idler}=0.562$~MHz and $N_\mathrm{signal}=0.507$~MHz signal photons detected using fiber-coupled single-photon counting modules (Excelitas SPCM-AQRH-14-FC) with a quantum efficiency of $60~\%$ at 810~nm operating at room temperature. 
Using a $\tau_c=2~$ns coincidence window we measured the typical rate of coincidences to be $N_c=0.138$~MHz. The second-order coherence of the signal path conditioned on idler detection was measured to be $g^{(2)}(0) = 0.006$,  and the source brightness was $7.55\pm0.01\times10^6$~pairs/s/mW. 
For experiments presented above the pump power was varied between 50 and 150$~\mu$W.

\subsubsection{Target Reflectivity and Noise Source}
Finite target reflectivity is realised using  neutral density filters. To calibrate the filter loss and thus target reflectivity $\xi$, we take the ratio of the total signal count-rate with (without) the filter after subtracting the detector background measured with the SPDC source blocked, using integration times of up to 1 second to provide sensitivity at the highest attenuation levels.

To simulate noise in the system, an LED is driven by a low-noise current driver (Koheron DRV300-A-10) to apply a controllable background count to the signal detector to provide a constant noise independent of target reflectivity or position. Use of a low-noise current driver is necessary to make a comparison between CI and QI detection protocols, as CI is extremely susceptible to any drift in noise level. For the active jamming presented in Figs. 4 and 5 a function generator is used to add modulation on top of a DC offset.

\subsection{Log-Likelihood Analysis Framework}

\subsubsection{Single-shot detection probabilities}
In order to evaluate the log-likelihood ratio, it is necessary to calculate the underlying probabilities of detection and coincidence events that occur within a single coincidence window $\tau_c$. These probabilities depend on the average photon number generated in the process of SPDC $\bar{n}$, target reflectivity $\xi$, the average background on the signal ($\bar{n}_\mathrm{bg}$) and idler ($\bar{n}_\mathrm{bg,I}$) detectors, detection efficiency of the signal ($\eta_S$) and idler detectors ($\eta_S$). The detection efficiencies account for all system loss and finite quantum efficiency in the limit of a perfect target reflectivity. Below we extend the analysis framework presented in Ref.~\cite{murchie23} to the regime of detection in the presence of a Poissonian background source, as used in this  experiment setup. 

To model the CI case, the probabilities of signal detector firing with the target present $p^\mathrm{CI}_{\mathrm{H1}}$ (or absent $p^\mathrm{CI}_{\mathrm{H0}}$) are given by
\begin{subequations}
\begin{align}
p^\mathrm{CI}_{\mathrm{H0}}&=1-\exp\left(-\bar{n}_\mathrm{bg}\eta_\mathrm{S}\right),\\
P^\mathrm{CI}_{\mathrm{H1}}&=1-\frac{1}{1+\gamma\eta_\mathrm{S}\xi\bar{n}}\exp\left(-\frac{\bar{n}_\mathrm{bg}\eta_\mathrm{S}}{(1+\eta_\mathrm{S}\gamma\xi\bar{n})}\right).
\end{align}\label{CI:click}
\end{subequations}

In the QI case, the single shot probabilities reflect the probability of a signal event within the coincidence time $\tau_c$ conditioned on the idler firing. When the object is absent, the detector only fires due to noise and the probability of a signal event is equivalent to the CI case with no target. With target present, it is necessary to derive the click probability by considering the idler conditioned signal state as a thermal-difference state\cite{Horoshko19}, resulting in the QI click probabilities given by
\begin{subequations}
\begin{align}
p^\mathrm{QI}_\mathrm{H0}&=p^\mathrm{CI}_\mathrm{H0},\\
p^\mathrm{QI}_\mathrm{H1}&=1-\frac{1}{p^\mathrm{QI}_\mathrm{I}}\biggl(\frac{1}{1+\bar{n}\xi\eta_\mathrm{S}\beta}\text{exp}\bigl(\eta_\mathrm{S}\bar{n}_\mathrm{bg}(\frac{\bar{n}\xi\eta_\mathrm{S}\beta}{1+\bar{n}\xi\eta_\mathrm{S}\beta}-1)\bigr) \nonumber \\
-&\frac{(1-p^\mathrm{QI}_\mathrm{I})}{1+\bar{n}_\mathrm{I:X}\xi\eta_\mathrm{S}\beta}\text{exp}\bigl(\eta_\mathrm{S}\bar{n}_\mathrm{bg}(\frac{\bar{n}_\mathrm{I:X}\xi\eta_\mathrm{S}\beta}{1+\bar{n}_\mathrm{I:X}\xi\eta_\mathrm{S}\beta}-1)\bigr)\biggr),
\end{align}\label{QI:click}
\end{subequations}
where $\bar{n}_\mathrm{I:X}=\bar{n}(1+\eta_I\bar{n}_\mathrm{bg,I}-\eta_\mathrm{I})/(1+\eta_\mathrm{I}\bar{n}_\mathrm{bg,I}+\bar{n}\eta_\mathrm{I})$ is the signal state mean photon number after conditioning from a no click event at the idler and $p^\mathrm{QI}_\mathrm{I}$ is the idler firing probability equal to
\begin{equation}
p^\mathrm{QI}_\mathrm{I}=1-\frac{1}{1+\eta_\mathrm{I}\bar{n}+\eta_\mathrm{I}\bar{n}_\mathrm{bg,I}}
\end{equation}

Furthermore, in these equations we introduce two additional parameters, $\gamma$ and $\beta$. These parameters allow for adjustments to match the data obtained for CI and QI respectively, to account for detector non-linearities and variations in the heralding efficiency due to changes in pump power and coincidence window duration.

\subsubsection{Log-Likelihood ratio}
In the limit of $k$ trials it is possible to express the LLV defined in Eq.~\ref{eq:LLV} in a linear form \cite{murchie23} dependent only upon the single-shot probabilities $p_{\mathrm{H0}}$ and $p_{\mathrm{H1}}$ using $\Lambda(x,k)=Mx+Ck$, where $M=\log((p_{\mathrm{H1}}*(1-p_{\mathrm{H0}}))/(p_{\mathrm{H0}}*(1-p_{\mathrm{H1}})))$ and $C=\log((1-p_{\mathrm{H1}})/(1-p_{\mathrm{H1}}))$.

For data acquired over an integration time $T$, for the CI LLV is calculated using $x$ as the number of detected signal events and $k=T/\tau_c$ corresponding to the number of trials of duration $\tau_c$ within the integration window. For the QI LLV, $x$ is the number of measured coincidence counts and $k$ is the number of idler firing events.

\subsubsection{Analytical Distinguishability}

To evaluate the analytical distinguishability it is necessary to calculate the mean and standard deviation of the underlying LLV distributions corresponding to target present and absent, and then integrate over this distribution to find the detection and false alarm probabilities from which $\phi$ is then calculated. 

In the limit where the underlying Poisson count distributions can be modelled as Gaussian \cite{murchie23} (valid for the data shown in the paper), the LLV distributions for H1 and H0 also become Gaussian, with mean value  $\mu_{Hi:\Lambda}=M\bar{x}_{Hi}+Ck$ and a standard deviation equal to $\sigma_{Hi:\Lambda}=M\sigma_{Hi}/\sqrt{N_\mathrm{av}}$, where $N_\mathrm{av}$ corresponds to the average window length being considered.

For the case of CI, as above $k=T/\tau_c$,  with mean signal levels given by $\bar{x}^\mathrm{CI}_{Hi}=kp_{Hi}^\mathrm{CI}$ and standard deviation $\sigma^\mathrm{CI}_{Hi}=\sqrt{kp^\mathrm{CI}_{Hi}(1-p^\mathrm{CI}_{Hi})}$. For the QI case, we use $k=(T/\tau_c)p^\mathrm{QI}_I$ corresponding to the average number of times the idler fires. The mean coincidence rate is then $\bar{x}^\mathrm{QI}_{Hi}=kp_{Hi}^\mathrm{QI}$ with standard deviation $\sigma^\mathrm{QI}_{Hi}=\sqrt{kp^\mathrm{QI}_{Hi}(1-p^\mathrm{QI}_{Hi})}$.

Defining a threshold LLV value $\Lambda_c$, the probabilities of detection and false alarm can then be evaluated as
\begin{subequations}
\begin{align}
P_\mathrm{D}&=\displaystyle\int_{\Lambda_c}^\infty \exp\left(-\frac{(x-\mu_{\mathrm{H1}:\Lambda})^2}{2\sigma_{\mathrm{H1}:\Lambda}^2}\right)dx,\\
P_\mathrm{FA}&=\displaystyle\int_{\Lambda_c}^\infty \exp\left(-\frac{(x-\mu_{\mathrm{H0}:\Lambda})^2}{2\sigma_{\mathrm{H0}:\Lambda}^2}\right)dx.
\end{align}
\end{subequations}
For the distinguishabilities quoted above we use threshold $\Lambda_c=0$ throughout the paper, with the ROC curve in Fig.~\ref{fig:ROC}\textbf{b} calculated by plotting $P_\mathrm{FA}$ versus $P_\mathrm{D}$ for different values of $\Lambda_c$.

\subsection{Estimating Experimental System Parameters}

To obtain parameters relevant for calculating single-shot probabilities using Eqs.~\ref{CI:click}-~\ref{QI:click}, the detection efficiencies $\eta_{S,I}$ are obtained using the known source brightness and comparing measured count rates at each detector in the absence of a simulated target, giving $\eta_{S,I}=0.2329,0.1958$ respectively for $\tau_c=2$~ns. For each scenario, background rates $\bar{n}_\mathrm{bg,S}$ and $\bar{n}_\mathrm{bg,I}$ are measured with the SPDC source blocked and only the noise source active. The emitted signal strength $\bar{n}$ is extracted from the idler detector with the pump active. This, combined with the calibrated target reflectivity $\xi$, provides the full set of system parameters. In the analysis above, a subset of data points (typically the first 100) for the target absent and present cases are compared against the theoretical model, using free parameters $\beta$ and $\gamma$ to fit against the measured single shot probabilities. Typically both parameters yield values close to unity corresponding to correcting only for the slight change of filter attenuation when replaced in the beam.

\subsection{Dynamic Background Tracking}

For the slow classical jamming data shown in Fig.~\ref{fig:jamming}\textbf{a} the system parameters are initially obtained by performing analysis of the first 200 measurements with static background $\bar{n}_\mathrm{bg,S}$. We then define 25 discrete background levels in the range 2.1-2.7 MHz, and for each value recalculate the single shot click probabilities in Eqs.~\ref{CI:click} and \ref{QI:click} using a re-scaled value of $\bar{n}_\mathrm{bg,S}'$. Subsequently the experimental data obtained with jamming engaged was analysed by calculating the LLV for each point in time by using the relevant click probabilities associated with the level that closest matches the instantaneous signal counts to dynamically track slow changes in background.

\section{Data Availability}
The datasets used in this work are available at [DOI to be added].

\section{References}

\section{Acknowledgements}
This project was funded by the UK Ministry of Defence.

\section{Author Contributions}
M.M. and J.P. devised the experiment; M.M. conducted the experiment, R.M. and J.J. developed the underpinning theoretical framework; J.P., R.M. and M.M. performed analysis of the data. All authors contributed to the discussion of results and preparation of the manuscript.

\section{Competing Interests}
The authors declare no competing interests.
\section{Additional information}
\subsection{Correspondence} Correspondence and requests for materials should be addressed to Jonathan D. Pritchard.

\end{document}